\documentclass{PoS}

\usepackage{booktabs}
\usepackage{multirow}
\usepackage{graphicx}
\usepackage{subcaption}
\usepackage{amsmath}
\usepackage{textcomp}

\title{MILC staggered conjugate gradient performance on Intel KNL}

\ShortTitle{MILC Staggered CG Performance on Intel KNL}

\author{\speaker{Ruizi Li}\\%\thanks{A footnote may follow.}\\
        Department of Physics, Indiana University, Bloomington IN 47405, USA\\
        E-mail: \email{ruizli@umail.iu.edu}}

\author{Carleton DeTar\\
        Department of Physics and Astronomy, University of Utah, Salt Lake City, Utah, 84112, USA\\
        E-mail: \email{detar@physics.utah.edu}}

\author{Douglas Doerfler\\
        National Energy Research Scientific Computing Center, Lawrence Berkeley National Laboratory, Berkeley, CA 94720, USA\\
        E-mail: \email{dwdoerf@lbl.gov}}

\author{Steven Gottlieb\\
        Department of Physics, Indiana University, Bloomington IN 47405, USA\\
        E-mail: \email{sg@indiana.edu}}

\author{Ashish Jha\\
        Software and Services Group, Intel Corporation, Hillsboro OR 97124, USA\\
        E-mail: \email{ashish.jha@intel.com}}

\author{Dhiraj Kalamkar\\
        Parallel Computing Lab, Intel Labs, Bangalore, India 560103\\
        E-mail: \email{dhiraj.d.kalamkar@intel.com}}
       
\author{Doug Toussaint\\
        Physics Department, University of Arizona, Tucson, AZ 85721, USA\\
        E-mail: \email{doug@physics.arizona.edu}}

\abstract{We review our work done to optimize the staggered conjugate gradient (CG) algorithm in the MILC code for use with the Intel Knights Landing (KNL) architecture. 
	KNL is the second generation Intel Xeon Phi processor. 
	It is capable of massive thread parallelism, data parallelism, and high on-board memory bandwidth 
	and is being adopted in supercomputing centers for scientific research. 
	The CG solver consumes the majority of time in production running, so we have spent most of our effort on it. 
	We compare performance of an MPI+OpenMP baseline version of the MILC code with a version incorporating the QPhiX staggered CG solver, 
	for both one-node and multi-node runs.}

\FullConference{34th annual International Symposium on Lattice Field Theory\\
		24-30 July 2016\\
		University of Southampton, UK}

\begin{document}

\section{Introduction}

In recent decades, lattice calculations have been performed on high-end supercomputers and clusters, 
looking for increased computing capability and capacity. 
The MILC collaboration has been using GPU clusters, 
listed among the Top 500, 
for boosting its code performance. 
Computers using Intel Xeon Phi processors, 
starting with the Knights Corner (KNC) coprocessor, 
and continuing with the current generation Knights Landing (KNL) processor, 
are also on the Top 500 list. 
We are porting the MILC code to the KNL processor and optimizing it 
through our participation in the Intel Parallel Computing Center (IPCC) Program, 
at the IPCC at Indiana University. 
We are also part of the NERSC Exascale Science Applications Program (NESAP). 
These programs provide us with access to KNL and other Intel Xeon products. \\
\\
This article is organized as follows. 
The second section gives a brief introduction to the KNL architecture. 
The third section describes the library we developed for the staggered conjugate gradient algorithm. 
The fourth section contains benchmarks and CG performance results, 
showing the performance improvement and comparing with several Intel architectures. 
The conclusions are in the final section.

\section{Intel Xeon Phi Knights Landing architecture}

Knights Landing is the second generation of the Intel Many Integrated Core (MIC) architecture 
and the first standalone processor in the Xeon Phi series. 
Its peak performance of over $3$ TFlop/s or $6$ TFlop/s in double or single precision (DP or SP) 
is over two times higher than that of the KNC. 
It is also more power efficient. 
In addition, it implements the new 512-bit AVX512 ISA and is also compatible with prior vector ISA's such as AVX2, AVX, and SSE.
It can be attached to the Omni-Path Fabric, 
which is an interconnection network recently developed by Intel. 
The network of the clusters we used, however, is Infiniband. 
Each node contains one chip, with $64$ to $72$ cores (from version 7210 to 7290) tiled in pairs. 
Each core has four threads and two 512-bit vector processing units (VPUs), 
as opposed to four threads and one VPU for KNC. 
The other significant feature is the in-package high bandwidth memory called MCDRAM with capacity of $8$ or $16$ GB. 
It offers up to $450$ GB/s stream bandwidth and $380$ GB/s read-only bandwidth. 
KNL can be configured in multiple clustering and memory modes. 
Memory modes can be Cache, Flat or Hybrid where MCDRAM is configured as memory-side cache (the Cache mode), 
or as memory (the Flat mode) or hybrid (cache and memory). 
Clustering modes include, for instance, quadrant and hemisphere\cite{ClusterModes}. 
The bus to off-board memory is a 6-channel DDR4 with up to $115$ GB/s bandwidth.

\section{Staggered QPhiX library}

\begin{figure}[ht]
\centering
\includegraphics[width=10cm]{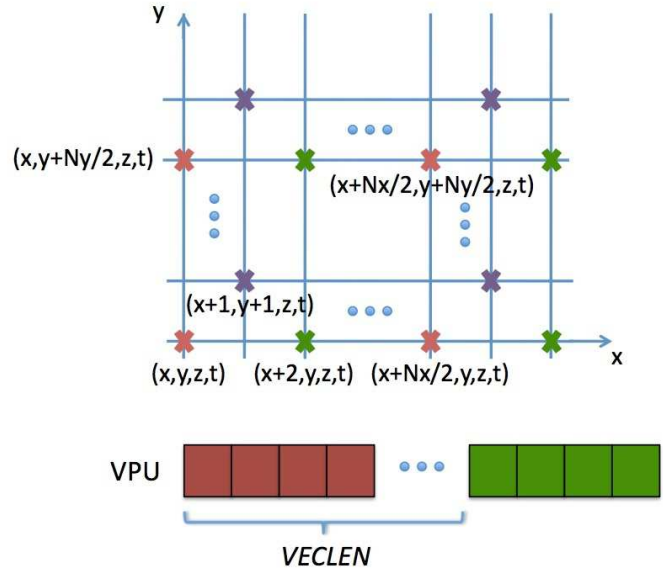}
\caption{\textit{Data layout showing two dimensions. 
The data is fused along both x and y directions. 
Data in the same color (red, green, purple) is stored contiguously in memory and cache.} }
\label{layout}
\end{figure}

The staggered QPhiX library was first developed for KNC\cite{Lattice2015}, 
adapted from the open-source QPhiX library for Wilson quarks\cite{wilsonQPhiX}\cite{wilsonQPhiXKNL}. 
It has been extended to KNL, 
including single-mass and multi-mass CG algorithms, 
and is being developed for other routines in the MILC evolution code. \\
\\
The library supports three Intel-architecture instructions, \textit{i.e.}, SSE, AVX2, AVX512. 
It includes a code generator that generates an intrinsic kernel file for each targeted algorithm or routine, 
for instance, staggered dslash.  
The other part of the library wraps kernel routines and contains higher-level algorithms, 
such as staggered multi-mass CG. 
It achieves both OpenMP and MPI parallelism. \\
\\
The top-level data layout is slightly different from that of the Wilson QPhiX library 
and is based on the Grid library\cite{Grid} developed by Peter Boyle.  
Data on the lattice is fused along three and four dimensions, 
with respect to double and single precision. 
In double precision, 
on a lattice of size $N_{x}, N_{y}, N_{z}, N_{t}$ along $x, y, z, t$ direction, 
data on sites $(x, y, z, t), (x+N_{x}/2, y, z, t), (x, y+N_{y}/2, z, t), ... ,$ $(x+N_{x}/2, y+N_{y}/2, z+N_{z}/2, t)$ 
is stored contiguously in memory, as shown in Figure~\ref{layout}. 
One advantage of this is simplified data fetching on the boundary, 
which reduces to just one vector permuting or swizzling intrinsic function. 

\noindent{Lattice data} is stored as arrays of structures of arrays, \textit{e.g.}, 

\begin{centering}
\begin{eqnarray*}
typedef &Real &*KS\_Color\_Vector[3][2][VECLEN]; \\[5pt]
typedef &Real &*Gauge[8][3][3][2][VECLEN]; \\[10pt]
\end{eqnarray*}
\end{centering}
%\newline
\noindent{where} $VECLEN = 8$ or $16$ for DP or SP, respectively, and $Real$ is $double$ or $float$. 
Gauge fields are stored 8-way, \textit{i.e.}, including both forward and backward links. 
This differs from the standard MILC gauge field structure, which stores forward links only by default. 
The 8-way gauge storage enhances data locality; 
thus we expect it to reduce the data access latency. 
The disadvantage, however, is that it doubles the gauge memory footprint during the staggered CG inversion step. 
A quantitative study of performance impact from two versions of the gauge storage has yet to be carried out.

\section{Benchmarks and results}

Our benchmarks focus on the double-precision staggered multi-mass CG algorithm. 
We optimize and benchmark this routine  
because it is the most time-consuming routine in the MILC evolution code, 
taking over $80\%$ of the time. 
As is well known, the staggered CG is memory-bandwidth bound, 
and staggered multi-mass CG is even more severely bandwidth bound. 
We show performance results in Gflop/s.  
Included in Table~1 is the list of machines we use for benchmarks. 
Clusters listed are the Intel Endeavor cluster and the NERSC Cori (Phase 1) cluster. 
They provide the Intel KNL, Broadwell, and Haswell architecture. \\

%configs table here ... 
\begin{table}[b]
\begin{center}
%\begin{small}
\begin{tabular}{|c|c|c|}
\hline\hline
	Cluster						&		Machine			&		Feature	\\
\hline\hline	
\multirow{9}{*}{Intel Endeavor Cluster}								&	\multirow{3}{*}{KNL 7210}	&	64 Cores @ $1.3$ GHz, \\
								&						&	8 or 16 GB MCDRAM, \\
								&						&	$6 \times 16$ GB DDR4 @ $2.1$ GHz \\
\cline{2-3}
								&	\multirow{3}{*}{KNL 7250}&	68 Cores @ $1.4$ GHz, \\
		&						&	8 or 16 GB MCDRAM, \\
								&						&	$6 \times 16$ GB DDR4 @ $2.4$ GHz \\
\cline{2-3}
								&	\multirow{3}{*}{Broadwell}					&	Dual Socket processor E5Z2697 v4, \\
								&			&	18 Cores/Socket, 36 Cores @ $2.3$ GHz, \\
								&						&	128 GB DDR4 @ $2.4 $ GHz \\
\hline\hline
\multirow{3}{*}{NERSC Cori Cluster}									&	\multirow{3}{*}{Haswell}					&	Dual Socket processor, \\
		&			&	16 cores/socket @ $2.3$ GHz, \\
								&						&	128 GB DDR4 @ $2.1$ GHz \\
\hline\hline								
\end{tabular}
\label{Configs}
\caption{Clusters used for the benchmarks.}
\end{center}
\end{table}

\noindent{We} use the application {\tt su3\_rhmc\_hisq} as a sample executable, 
a staggered-fermion rational hybrid Monte Carlo evolution code, 
to compare multi-mass CG performance across baseline MILC code with both MPI and hybrid MPI+OpenMP, 
and MILC with QPhiX optimization and hybrid MPI+OpenMP. 
The number of quark masses is set around nine. \\

%\centering
\begin{figure}[ht]
\begin{subfigure}{.5\textwidth}
\centering
\includegraphics[width=6.5cm]{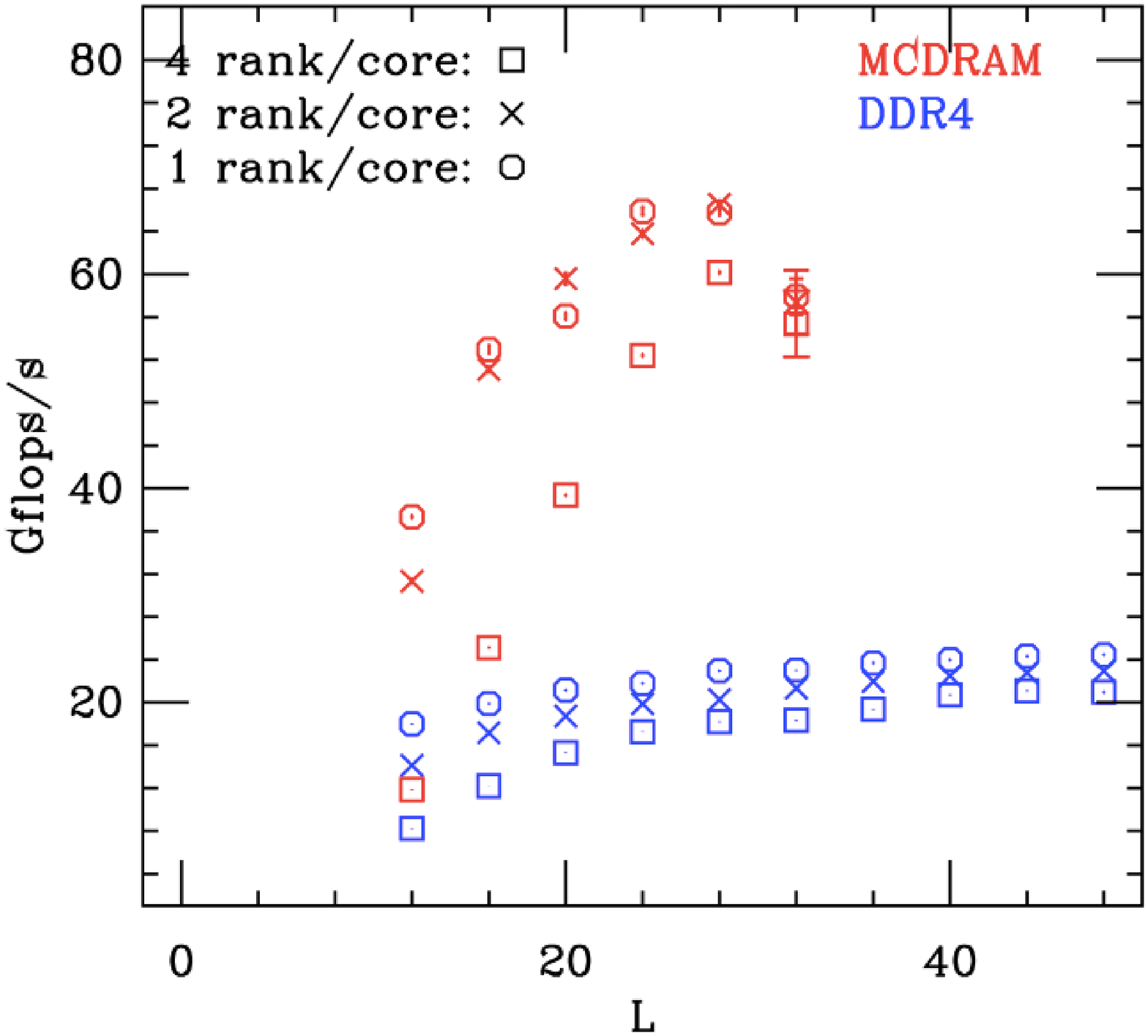}
\label{MILCKNL1}
\end{subfigure}
\begin{subfigure}{.5\textwidth}
\centering
\includegraphics[width=6.2cm]{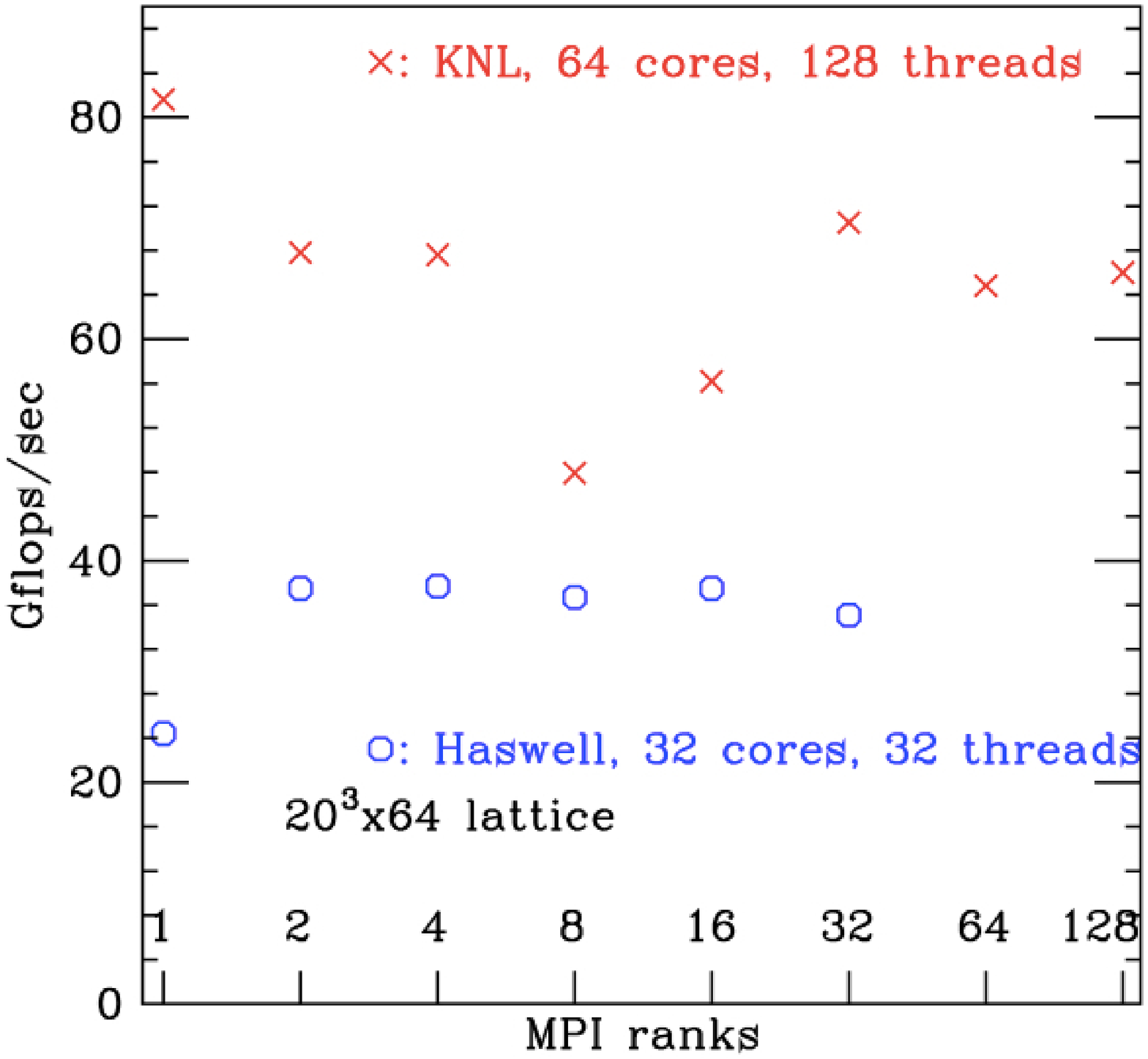}
\label{MILCKNL2}
\end{subfigure}
\caption{ \textit{Baseline single-node benchmark performance with MPI (left) and MPI+OpenMP (right). 
The vertical axis shows the total performance.  
In the left plot the horizontal axis marks the lattice size $L$ along each dimension, 
and data in red and blue give the performance w/ and wo/ MCDRAM usage. 
Here the total number of MPI ranks is 64, 128, and 256, 
and the number of cores used is 64. 
In the right plot the horizontal axis marks the number of MPI ranks, 
and data in red and blue give the performance on KNL and Haswell. 
The total number of MPI ranks here is 128 and 32, 
and the number of cores used is 64 and 32 on KNL and Haswell, respectively.} }
\label{MILCKNL}
\end{figure}

\begin{figure}[ht]
\centering
\includegraphics[width=15.5cm]{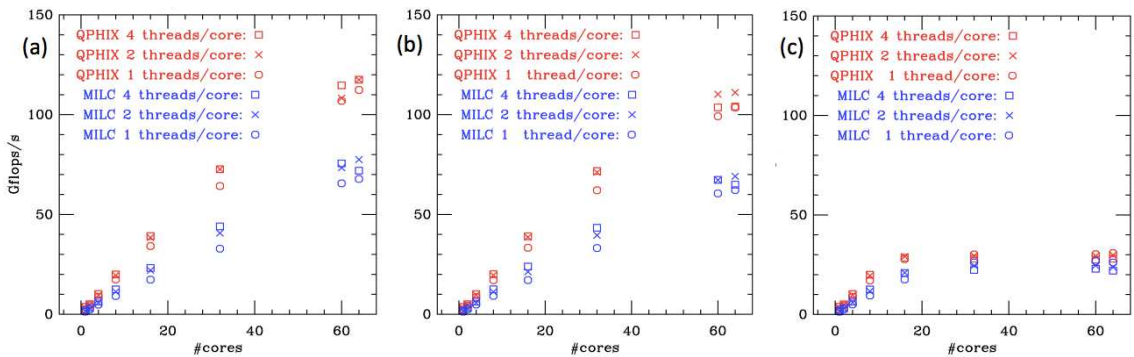}
\caption{\textit{Optimized vs.\ baseline code weak scaling benchmark on one KNL 7250. 
The vertical axis shows the total performance, 
and the horizontal axis shows the number of cores. 
Plots (a), (b), and (c) show the performance in the Flat mode using MCDRAM, the Cache mode, 
and the Flat mode without MCDRAM usage, respectively.} }
\label{QPhiXvsMILC1}
\end{figure}

\begin{figure}[ht]
\centering
\includegraphics[width=14.5cm]{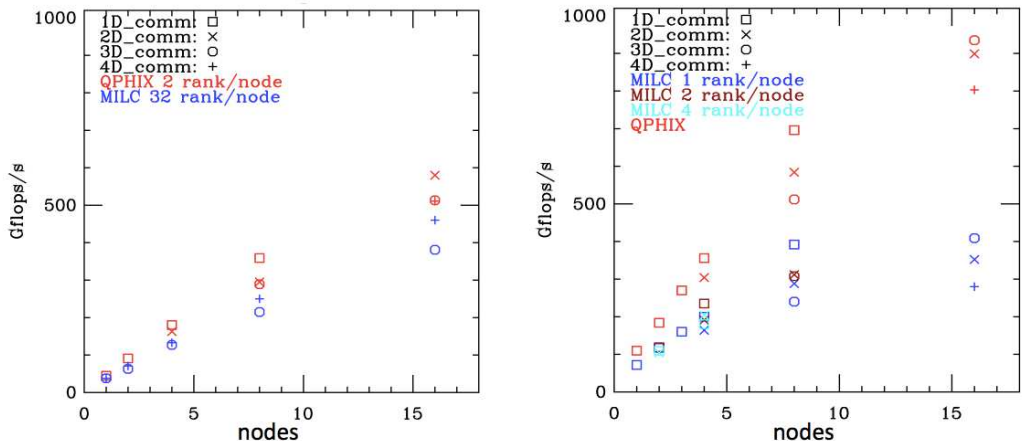}
\caption{\textit{Optimized (QPhiX) vs.\ baseline (MILC) code weak scaling benchmark on Broadwell (left) and KNL 7210 (right) up to 16 nodes. 
The vertical axis shows the total performance in Gflop/s, 
and the horizontal axis shows the number of nodes. 
The number of cores used per node is 60 on KNL and 32 on Broadwell. 
The number of MPI ranks on KNL with QPhiX is one per node. 
Lattice size is  $24^3 \times 60$ per node on both machines.} }
\label{QPhiXvsMILC2}
\end{figure}

\noindent{Figure}~\ref{MILCKNL} (L) shows the performance of the baseline MILC code with MPI on various lattice sizes $L^4$. 
The run is on one KNL 7250 node using $64$ cores. 
Runs use as many as 256 MPI ranks, \textit{i.e.}, four ranks per core. 
As expected, using MCDRAM increases the performance up to three times. 
The performance has a sweet spot at $L=24$ or $28$, 
with a peak value of around $65$ Gflop/s. 
MCDRAM is used either in the Cache or Flat mode, 
and in our study those choices lead to less than $10\%$ performance difference,  
with the Flat mode giving a slightly higher value. 
Thus, they are not shown separately. 
The clustering mode is all-to-all, 
as used in most of our benchmarks. 
Comparing different clustering modes has not been a significant performance tuning effort yet. 
We also used quadrant clustering mode in some of our benchmarks, 
and have observed the performance similar to all-to-all, 
which is consistent with the Wilson CG case\cite{wilsonQPhiXKNL}. 
Further study of clustering mode effects is to be undertaken. \\
\\
To compare the efficiency of MPI and OpenMP parallelization on KNL, 
we carry out another set of single-node runs. 
We fix the problem size at $20^3 \times 64$, 
and the total amount of hardware resources, 
or equivalently total number of cores and OpenMP threads, 
while varying the number of MPI ranks. 
The performance is shown in the right of Figure~\ref{MILCKNL}.  
These tasks compare KNL and Haswell, 
in which the performance on KNL peaks at one MPI rank, 
while on Haswell, at two or more MPI ranks. 
This is consistent with the fact that a Cori Phase 1 node contains two Haswell chips while a KNL node has only one chip. 
The performance fluctuation with various MPI ranks on KNL is much more severe than on Haswell. 
We expect this to be caused by load imbalance, 
though we are searching for a definite explanation. 
The number of threads per core is set so each machine gives its best performance, 
thus providing a fair comparison of the architectures. 
We see the best performance on KNL is around twice that of Haswell. \\
\\
We observe from benchmarks of the baseline MILC code 
that MCDRAM is the key point for boosting the performance on KNL, 
whichever way we use it, 
and OpenMP seems to work a bit better than MPI, 
at least on one KNL chip. 
We still find these to hold after optimizing the code with QPhiX. \\
\\
The performance of QPhiX staggered dslash increases with increased lattice size. 
The routine's model bandwidth is calculated as the least amount of data being fetched from memory, 
excluding repeated fetching due to cache misses, 
and is up to $80\%$ of the peak read bandwidth with hardware prefetches only. 
The Intel VTune performance analysis software reports around $15\%$ cache misses in this routine. \\
\\
We compare performance before and after QPhiX optimization 
on one node in Figure~\ref{QPhiXvsMILC1}, 
and multiple nodes in Figure~\ref{QPhiXvsMILC2}. 
Figure~\ref{QPhiXvsMILC1} shows the weak scaling benchmark performance up to 64 cores on a KNL 7250. 
Plots (a) and (b) use MCDRAM. 
Both the Flat mode (a) and the Cache mode (b) give similar performance. 
On KNL, QPhiX performs best with one MPI rank per node on one and
multiple nodes, which is the parallelization configuration here and
later in the multi-node benchmark.
Overall QPhiX gains 1.50x in performance over baseline MILC code. 
Plot (c) shows the performance in the Flat mode without MCDRAM. 
As expected, performance saturates quickly with an increased number of cores, 
due to the limited DDR4 bandwidth. \\
\\
Figure~\ref{QPhiXvsMILC2} compares the weak scaling performance of two versions of the code on KNL and Broadwell, 
with up to 16 nodes and up to 4-dimensional communications.  
The interconnect network in both cases is Infiniband. 
Note that in this, and some other benchmarks, 
the number of threads per core on KNL is set to be two, 
which gives the best overall performance including multiple nodes. 
Scaling on multiple nodes up to 16 is optimized further, 
compared to single node. 
The benefit comes from the non-blocking, one-time communication strategy, 
the same as in the Wilson QPhiX library. 
On the 16-node KNL cluster, QPhiX delivers 900 GFlop/s. 
This represents a 2.20x performance gain over baseline MILC. 
On the 16-node Broadwell cluster, the QPhiX gain is less significant. 
\\

%QPhiXvsMILC1 here ...

%QPhiXvsMILC2 here ...

\section{Conclusions}

We optimize and benchmark the staggered multi-mass CG algorithm, 
developing the staggered QPhiX library from the Wilson QPhiX library. 
We observe a performance improvement of around 1.50x on one KNL chip, 
and 2.20x on multiple nodes up to 16. 
Future work includes further improving the CG performance in staggered QPhiX, 
enabling software prefetch tuning, exploring various clustering modes, 
and optimizing other routines and algorithms in the MILC code, 
for instance, the gauge force and the fermion force calculations. \\
\\
\textbf{Acknowledgments:} 
Many thanks to B\'alint Jo\'o for great help and valuable discussions on
building and developing of the staggered QPhiX library. 
This work was supported in part by U.S. DOE under grants 
DE-SC0010120 (S.G.), 
DE-FG02-13ER41976 (D.T.) 
and the U.S. NSF under grant 
PHY10-034278 (C.D.). 
R.L.~and S.G.~thank Intel\textsuperscript{\textregistered} for its support of the Intel Parallel Computing Center at Indiana University.


\begin{thebibliography}{99}
\bibitem{ClusterModes} Colfax research website, http://colfaxresearch.com/knl-numa/
\bibitem{Lattice2015} R.~Li, S.~Gottlieb, \textit{Staggered Dslash Performance on Intel Xeon Phi Architecture}, PoS LATTICE \textbf{2014}, 034 (2015) [arXiv:1411.2087 [hep-lat]]. 
\bibitem{wilsonQPhiX} B.~Jo\'o {\it et al.}, 
   ISC 2013,
   Lecture Notes in Computer Science, Vol.~7905, 40 (2013), 
   J.M~Kunkel, T.~Ludwid, and WH.W. Meuer (Eds.).
\bibitem{wilsonQPhiXKNL} J.~Jeffers \textit{et al.}, \textit{Intel Xeon Phi Processor High Performance Programming Knights Landing Edition}, Chap.~26, ISBN: 978-0-12-809194-4.  
\bibitem{Grid} P.~Boyle \textit{et al.}, \textit{Grid: A next generation data parallel C++ QCD library}, PoS LATTICE \textbf{2015}, 023 (2015) [arXiv:1512.03487 [hep-lat]].
\end{thebibliography}
\end{document}